\documentclass[9pt, twocolumn, aps,jap]{article}

\usepackage{ulem}
\usepackage{amsmath}
\usepackage{lineno}
\newcommand\rr{\mathbf{r}}

\usepackage[dvipsnames]{xcolor}
\usepackage{graphicx}

\usepackage{geometry}
\usepackage{authblk}

\title{Imaging through a square multimode fiber by scanning focused spots with the memory effect}

\author[1,2]{Sylvain Mezil}
\author[1,*]{Ir\`ene Wang}
\author[1]{Emmanuel Bossy}

\affil[1]{Univ. Grenoble Alpes, CNRS, LIPhy, F-38000 Grenoble, France}
\affil[2]{Institut Langevin, ESPCI Paris, Universit\'e PSL, CNRS, 75005 Paris, France}

\affil[*]{Corresponding author: irene.wang@univ-grenoble-alpes.fr}
\date{\vspace{-1cm}}

\begin{document}

\twocolumn[
  \begin{@twocolumnfalse}
    \maketitle
    \begin{abstract}
      The existence of a shift-shift memory effect, whereby any translation of the input field induces translations in the output field in four symmetrical directions, has been observed in square waveguides by correlation measurements. Here we demonstrate that this memory effect is also observed in real space and can be put to use for imaging purposes. First, a focus is created at the output of a square-core multimode fiber, by wavefront shaping based on feedback from a guide-star. Then, thanks to the memory effect, four symmetrical spots can be scanned at the fiber output by shifting the wavefront at the fiber input.  We demonstrate that this property can be exploited to perform fluorescence imaging through the multimode fiber, without requiring the measurement of a transmission matrix.
\vspace{0.5cm}
    \end{abstract}
  \end{@twocolumnfalse}
]

Imaging at the tip of an endoscopic probe allows to see deeply, albeit invasively, into living tissues and organisms. In order to reduce the damage caused by its insertion, minimizing the diameter of endoscopic probes is an important issue. Multimode fibers (MMFs) present an advantageous ratio between the information content they carry and their footprint. However, optical signals transmitted by MMFs are scrambled by modal dispersion. To recover this information, most methods are based on prior knowledge of the scrambling process by either measuring the transmission matrix~\cite{choi2012scanner,vcivzmar2012exploiting,Loterie2015,Ohayon2018,Turtaev2018,VasquezLopez2018} or learning the output speckle patterns~\cite{Mahalati2013,Amitonova2018,Rahmani2018}. Whatever the calibration method, access to both the proximal and the distal end of the fiber is required.  
This, combined with the sensitivity of the calibration results to small perturbations of the fiber (deformations, temperature change), limits the potential applications of these methods.

By many aspects MMFs bear similarities with scattering layers: as light propagates through such layers, spatial information are also distorted or scrambled. To look inside a scattering environment, a range of calibration-free methods have emerged, based on the optical memory effect: in spite of multiple scattering, a transformation of the input field can be transmitted by the medium, resulting in the same transformation of the output field~\cite{feng1988correlations,judkewitz2015translation}. In this case, even if the output field is unknown, its correlations can be exploited for imaging~\cite{bertolotti2012non,katz2014non,Schott2015,Wang2021,Zhu2022}. Although memory-effect-based imaging has been reported in fiber bundles~\cite{porat2016widefield}, similar methods have not been extended to single MMFs, due to their general lack of a proper memory effect. A rotational memory effect has been observed in circular MMFs~\cite{amitonova2015rotational}, but it alone cannot allow 2D imaging. When combining the feedback signal from a guide-star with memory-effect considerations, it has been shown that enough information on the fiber transmission can be gathered, so that a small region at the fiber output can be imaged, using measurements from only one end~\cite{Li2021}. However, due to the absence of a real radial memory effect, the size and shape of this region varies with the location of the guide-star and reduces to a single point at the center of the fiber.

Recently, a special type of shift-shift memory effect in square-core multimode fibers (SqMMFs) has been observed~\cite{CaravacaAguirre2021}. It has the ability to produce two-dimensional (2D) translations of the output pattern, as opposed to the sole azimuthal rotation reported in circular MMFs. According to this effect, that we will now systematically refer to as the memory effect, any field at the input of a SqMMF results in four speckle patterns shifting along four symmetrical directions at its output. These four directions are mirror reflections of the input shift direction relative to the edges of the square core. This memory effect was demonstrated by cross-correlating speckle patterns in the output plane for different shifts induced in the input plane, producing four peaks in the cross-correlation space that correspond to four speckle patterns shifting in the output plane. 

The aim of the present work is to exploit this memory effect to perform fluorescence imaging in SqMMFs by scanning focused spots in the sample plane. In a parallel work, we also recently demonstrated that the memory effect in sqMMFs can be also exploited for imaging without any prior assistance, through a correlation-imaging approach with speckle patterns in the sample plane~\cite{bouchet2022speckle}. This however requires both significant averaging over speckle realizations and solving an ill-posed inverse problem. The method proposed here has the advantage to provide a straightforward reconstruction from a single scan, directly through a focus-scanning microscopy approach. Starting from a focused spot that we first obtain by use of a guide-star, we indeed demonstrate that the memory effect allows to simply scan four foci across the sample plane, as opposed to the previous observation that reported shifting peaks in the correlation domain~\cite{CaravacaAguirre2021,bouchet2022speckle}. These spots can be scanned around the guide-star generated focus,  over what we will call an isoplanetic patch, defined by the observed  range of the memory effect, typically $\sim$10~\textmu in our experiment. By exploiting the memory effect in SqMMFs to scan the focused spots in a sample plane, we experimentally demonstrate imaging of fluorescence bead.

Our proposed approach consists of two steps: first, the field at the proximal fiber facet is optimized to generate a focus at the fiber output using as feedback the intensity at a single point ('guide-star'). Then, the optimized input field is translated in 2D in the transverse plane, which induces the scan of four foci over the sample, placed at the distal facet, and enables the acquisition of a fluorescence image using a bucket detector  on the proximal side.

\begin{figure}[htb]
\centering
\includegraphics[width=1\linewidth]{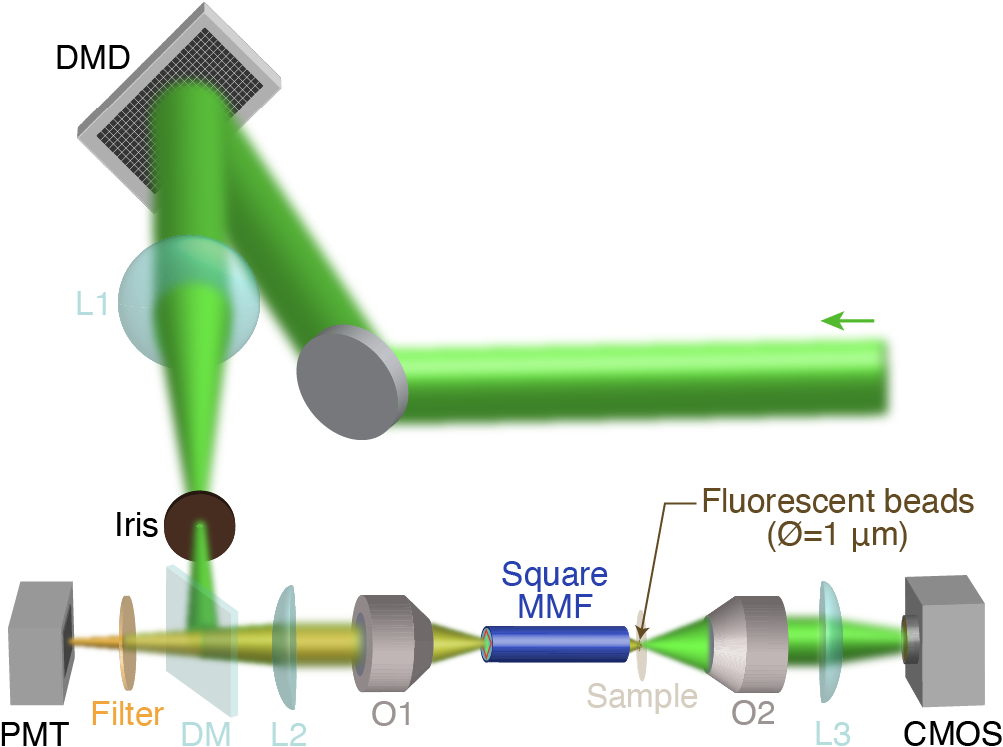}
\caption{Experimental setup. A 532-nm laser beam is sent onto a DMD. The pattern displayed on the DMD diffracts a phase-modulated field in its first order, which is selected by the iris, reflected by a dichroic mirror (DM) and imaged by a lens and an objective (O1) onto the fiber input facet. The fluorescent sample is placed at the fiber output; its emission is collected back through the fiber, transmitted by the dichroic mirror and detected by a photomultiplier (PMT). A camera (CMOS) at the fiber distal end provides an artificial guide-star (using the intensity of only one pixel) for wavefront shaping and transmission images of the sample.}
\label{fig:manip}
\end{figure}

Based on symmetry arguments in the case of an ideal square waveguide with mirror-like walls, we can show that, translating an optimized input field by $\Delta \rr_a=(x_a,y_a)$, leads, at the fiber output, to splitting of the initial focus into four spots, each of them moving in symmetrical directions: $\Delta \rr_a\!=\!(x_a,y_a)$, $\Delta \rr_a^\pm=(x_a,-y_a)$, $\Delta \rr_a^\mp=(-x_a,y_a)$ and $\Delta \rr_a^-=(-x_a,-y_a)$. 
We experimentally demonstrate this property using a setup for wavefront optimization and fluorescence imaging, as depicted in Fig.~\ref{fig:manip}.

\begin{figure}[ht!]
\includegraphics[width=1\linewidth]{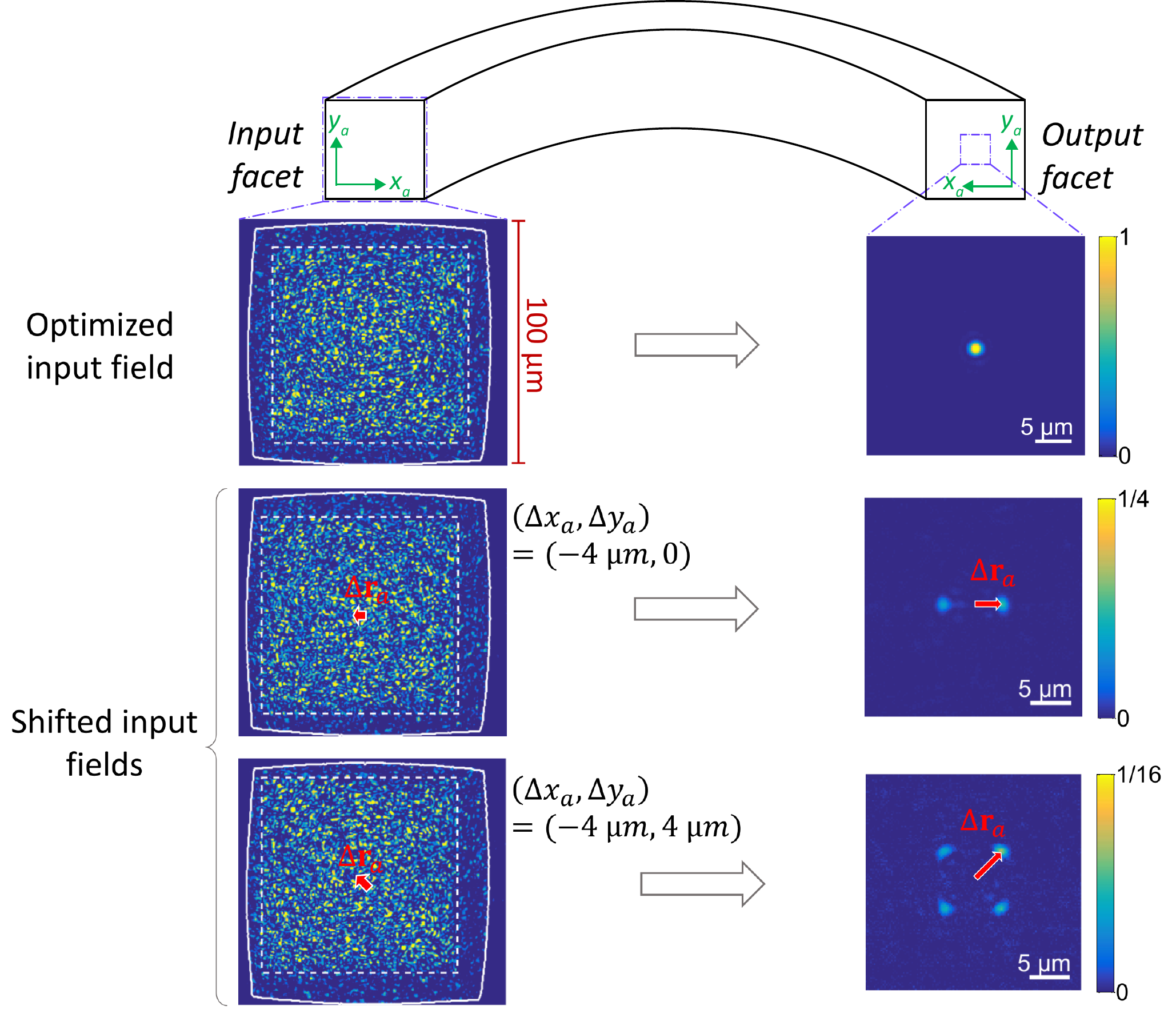}
\caption{Memory-effect induced scanning through a square multimode fiber. The left column shows camera images of the input facet of the fiber (the fiber core is delimited by a \textit{solid white line}) and the right column the corresponding images of the fiber output. After optimisation of the input wavefront in a 80$\times$80~$\mu$m region in the fiber core (\textit{dotted write line}), the fiber output is a focused spot (\textit{top row}). By shifting the input field by 4~$\mu$m in the $x$ direction, the output becomes two spots displaced in opposite directions (\textit{middle}). When the input field is shifted in both $x$ and $y$ directions, four spots are translated in symmetrical directions (\textit{bottom}). Red arrows show the translation vector of the input field, $\Delta \rr_a$, in both image spaces.}
\label{fig:spots}
\end{figure}

The wavefront at the input of a SqMMF (CeramOptec, 100$\times$100~\textmu m$^2$ section and 30-mm long) is shaped using a DMD (Digital Micromirror Device) that displays binary Lee holograms~\cite{Lee1978}. Input modes consist of plane waves in the DMD plane, corresponding to focused spots at the fiber entrance. They have a spacing of 1~µm and cover a 80$\times$80~\textmu m$^2$ square centered on the fiber facet (6400 input modes), the 10~\textmu m margin on each side allowing the input pattern to translate while remaining inside the fiber core. To focus light at the fiber output, the relative phases of input modes can then be optimized using the feedback signal from a guide-star collected on any kind of bucket detector, provided here by the intensity measured at one single pixel of the camera plane conjugated to the fiber output facet. Using the DMD, the phase of each input mode is varied relative to that of a reference spot (located in the corner of the fiber facet). Five levels of phase, spanning the [0,2$\pi$] range, are sequentially acquired, while monitoring the feedback signal; the optimal phase for each mode is determined using a sinusoidal fit. Finally, the optimized wavefront is obtained by summing all input modes, each with its optimal phase. This leads to a single focus at the fiber output (Fig.~\ref{fig:spots}). 
The measured enhancement factor (computed as the ratio of the peak intensity of the focus divided by the average intensity of the background speckle over the remaining area of the fiber~\cite{Turtaev2018}) exceeds 2000.

After optimization, we shift the pattern in the input plane of the fiber by adding a tilt to the optimal phase map on the DMD. The input scan range is typically 20$\times$20~\textmu m$^2$ with a step of 0.5~\textmu m to 0.25~\textmu m in either direction. 

\begin{figure}[ht!]
\centering
\includegraphics[scale=0.9]{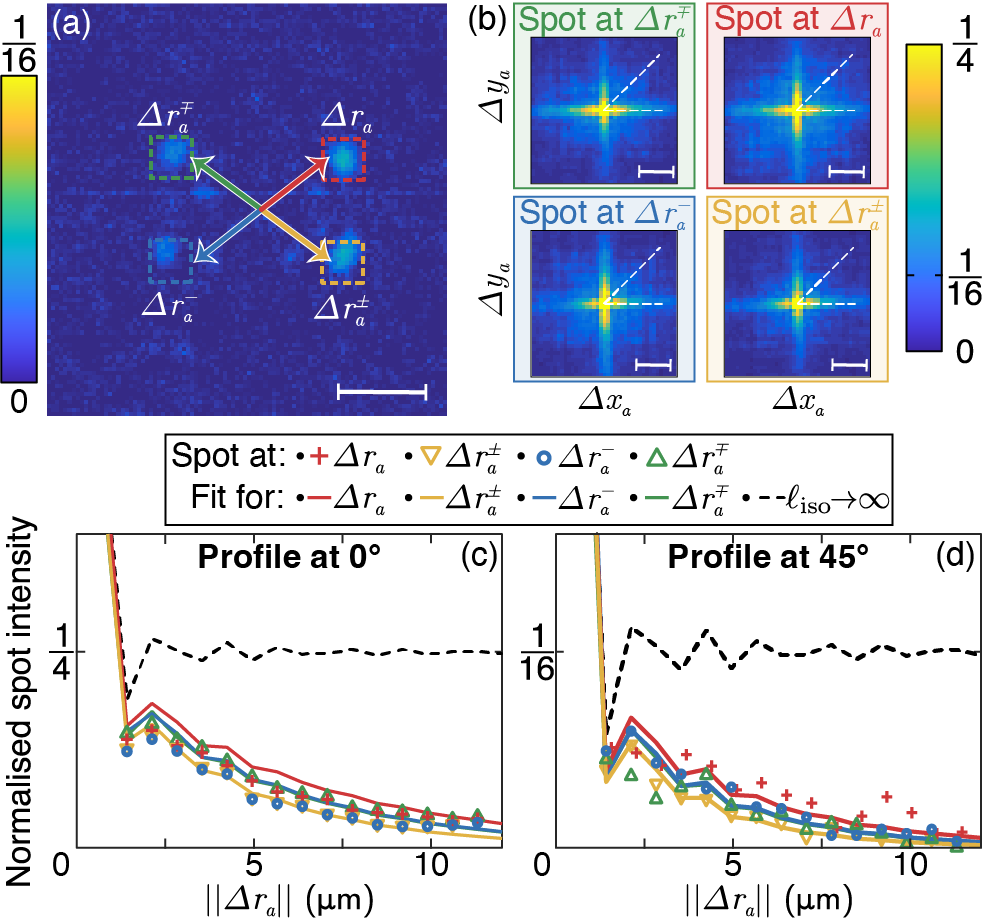}
\caption{Extension of the isoplanetic patch. 
The maximum value of the four peaks in the four directions $\Delta {r_a}^{\{\;\;, -, \pm, \mp \}}$ 
are extracted for each input shift (a), producing a map for each peak showing its intensity as a function of the input shift ($\Delta x_a$,$\Delta y_a$) (b). Line profiles at 0$^\circ$, (c) and at 45$^\circ$~(d) are extracted from these maps. Experimental values (\textit{symbols}), corresponding fits (\textit{colored lines}) and the case of an infinite memory effect (\textit{dashed line}) are plotted.
Estimated ranges for the memory effect $\ell_\text{ME}$ are 8.1, 6.7, 5.6 and 6.8~\textmu m for the spots at $\Delta \rr_a$, $\Delta \rr_a^{\mp}$, $\Delta \rr_a^{\pm}$,$\Delta \rr_a^{-}$, respectively. All intensities values are normalized by the optimized spot maximum intensity. Scale bar: 5~\textmu m.}
\label{fig:extent}
\end{figure}

We first verified that foci are indeed scanned at the fiber output by acquiring a camera image of the output intensity for each input shift. Fig.~\ref{fig:spots} shows that, when the input shift direction is along an edge of the square core, the focus splits into two spots, while, in the general case, it splits into four spots moving in symmetrical directions. These results confirms the existence of the memory effect directly from the translation of a focal spot, which was to date only demonstrated via cross-correlation of speckle patterns~\cite{CaravacaAguirre2021,bouchet2022speckle}. The intensities of these spots, normalized by the initial focus intensity, are lower than the values predicted for the ideal case ($1/4$ for the two-spots case and $1/16$ for the four-spots case, although the ratio between two and four spots is preserved. Moreover, we observe approximately equal intensities on the four spots, as expected.

The spots intensities, as a function of the input translation distance, are presented on Fig.~\ref{fig:extent}, where the maximum intensity of each peak is extracted from the images and displayed on 2D maps as a function of input shift $(\Delta x_a, \Delta y_a)$ (Fig.~\ref{fig:extent}(b)). As predicted, higher intensity values are observed for shift directions along the fiber core edge (horizontal and vertical). Line profiles extracted from these maps show that the measured intensities approach the ideal case at very small input shifts, but they quickly drop from $||\Delta \rr_a||\sim 3$~\textmu m, and finally vanish around $||\Delta \rr_a||\sim 10$~\textmu m. This maximum translation distance is in agreement with the extent of the memory effect observed in Ref.~\cite{CaravacaAguirre2021} from computed correlations.

To quantify more precisely the extent of the isoplanetic patch, we fitted these 2D maps by assuming four diffraction-limited Airy spots (at their theoretical positions), which intensity decreases exponentially as a function of $||\Delta \rr_a||$ with a characteristic length $\ell_\text{ME}$, which is hereafter referred to as the range of the memory effect.
This empirical expression describes reasonably well the experimental measurements, as shown on the line profiles of Fig.~\ref{fig:extent}(c) and (d), and the fits yield $\ell_\text{ME}$ values ranging between 5 and 8~\textmu m.

By measuring $\ell_\text{ME}$ for different guide-star location on the fiber output facet, we observed that $\ell_\text{ME}$ remains mostly unchanged within the fiber core, and is reduced only around the corners. Interestingly, unlike circular fibers in which case the isoplanetic patch is highly elongated at the fiber edge and reduces to a point at the center~\cite{Li2021}, SqMMFs exhibit a much more stable behaviour across the whole fiber facet,  although a maximal $\ell_\text{ME}$ is observed at the center.

\begin{figure}[t!]
\centering
\includegraphics[scale=0.95]{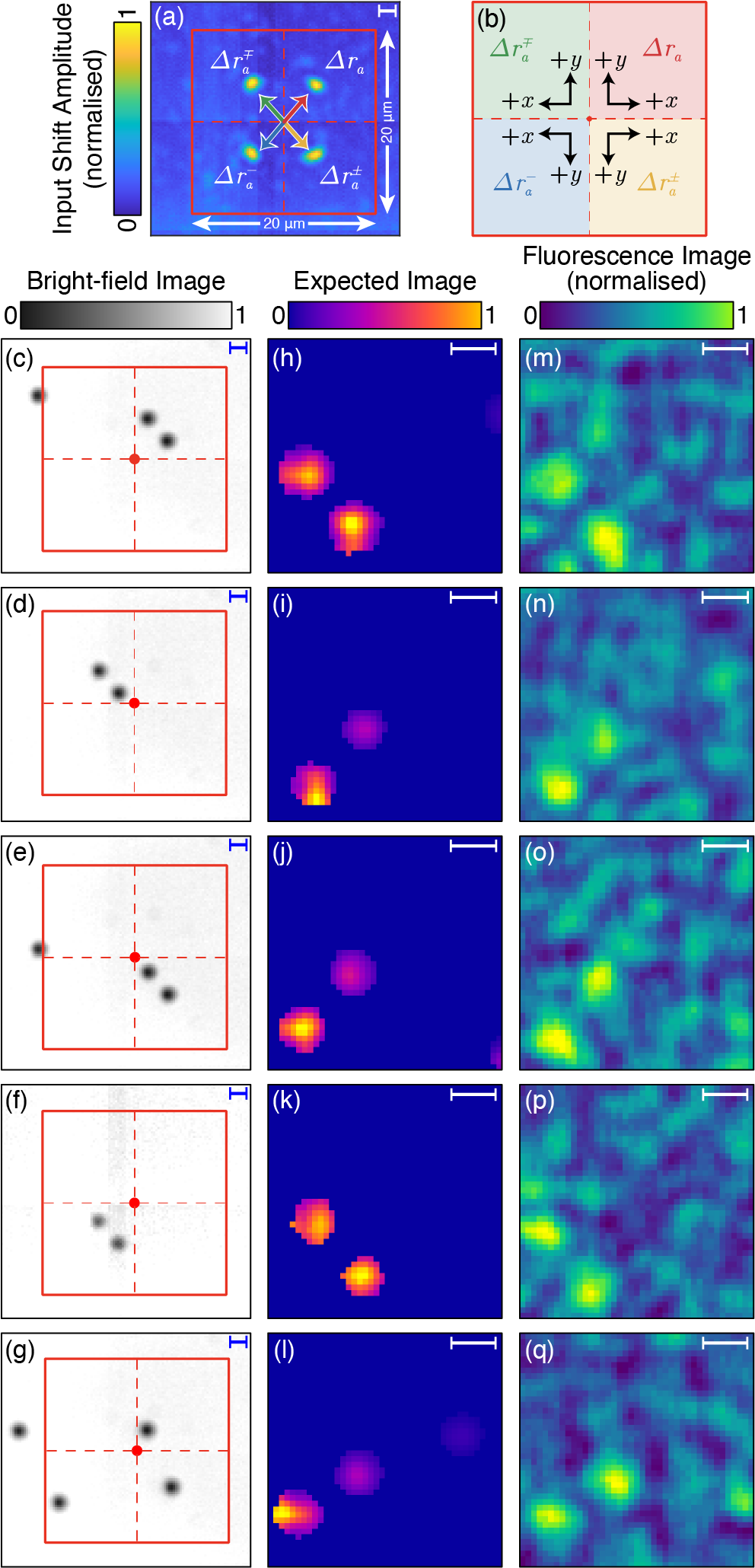}
\caption{Imaging results. (a)~Acquisition of the output facet for a shift of $\Delta_x$=$\Delta_y$=3~\textmu m (from  the case in (c,h,m)). (b)~Representation of the scan direction of each quadrant associated with $\Delta {r_a}^{\{\;\;, -, \pm, \mp \}}$.  
(c-g)~Bright field image of the sample. Red square delimits the scanned area. Dashed lines separate the quadrants of each focus point. (h-l)~Reconstruction of the expected image. 
(m-q)~Maximum of the fluorescent signal acquired on the PMT. Cases are associated per line (i.e., (c)-(h)-(m)). Scale bar: 2~\textmu m.}
\label{fig:results}
\end{figure}

This memory effect was then exploited for fluorescence imaging by placing a coverslip with dispersed 1~\textmu m-diameter fluorescent beads at the fiber output. Results are displayed in Fig.~\ref{fig:results}. The first column of Fig.~\ref{fig:results} (c-g) displays the bright-field images of the sample. Each scanned area contains 2 to 3 fluorescent beads; cases (c)-(f) having two beads in only one quadrant, while case (g) have three beads in three different ones.
For this experiment, the optimised input field is shifted from 0 to 10~\textmu m (in $x$ and $y$~directions) so each output spot 
scans one quadrant of the sample (each spot scans a 10x10~\textmu m$^2$ area, making the total scanned area of 20x20~\textmu m$^2$). The four quadrants and the associated spots are represented in Fig.~\ref{fig:results}(a-b). We chose to represent the fluorescence image in the input translation frame $\Delta \rr_a$. The resulting 10x10~\textmu m$^2$ image is hence the superposition of the four quadrants that are simultaneously scanned. Since the scan direction of the spot differs in each quadrant (see axes in Fig.~\ref{fig:results}(b)), each contribution provides a flipped image of the real sample.

The second column of Fig.~\ref{fig:results} (i.e., (h)-(l)) corresponds to the expected results: they are obtained by folding the bright-field images and take into account the exponential decay of the spots intensities. The center of the scan is determined from images of the output spots acquired during the scan. For the cases (c)-(f), the sample is moved so that the same two beads are successively in each quadrant of the scanned area. In the case of Fig.~\ref{fig:results}(c) where the beads are scanned by the spot associated to $\Delta \rr_a$, the expected image is in the same direction as the bright-field image.
The one from the second case (Fig.~\ref{fig:results}(d), associated with $\Delta \rr_a^\mp$) preserves the vertical axis but flips the horizontal one, while the reverse is observed 
from the third case (Fig.~\ref{fig:results}(e), associated with $\Delta \rr_a^\pm$)). 
Finally, the one from the  fourth case (Fig.~\ref{fig:results}(g), associated with $\Delta \rr_a^-$)) flips both axis. (The fifth case will be discussed later.) Furthermore, due to the limited memory effect (see Fig.~\ref{fig:extent}), the expected beads intensities  decrease with distance from the center. 

The third column (Fig.~\ref{fig:results}(m-q)) corresponds to the images reconstructed from the photomultiplier (PMT). One can see the very good agreement between the expected images and the actual reconstructed ones. For the cases  Fig.~\ref{fig:results}(m-p), the two beads are retrieved 
in the expected position. A small shift can be observed for the fourth case, which is attributed to a shift of the sample during acquisition. A background signal is also observed, due to the fraction of light that is not focused at the fiber output and forms a background speckle (part of this speckle can be observed in Fig.~\ref{fig:results}(a)). This speckle illuminates the sample on a 100x100~\textmu m$^2$ area (the fiber core) where several other beads are present.  However, while this background decreases the signal-to-noise ratio, it clearly remains possible to detect and the localize of the fluorescent beads.

Since the signal detected by the PMT is the fluorescence excited simultaneously by the four foci, it is impossible to retrieve from which quadrant the signal originates from a single image.  As long as all the beads are within one quadrant, this ambiguity affects only the direction of the image (which can undergo mirror symmetries). Conversely, if several beads are in different quadrants, this can lead to a wrong image reconstruction as illustrated in  Fig.~\ref{fig:results}(g,l,q). 
Here, there are three beads dispersed in the upper-right, in the lower-right and, further, in the lower-left quadrants (associated with $\Delta \rr_a$, $\Delta \rr_a^\pm$ and $\Delta \rr_a^-$, respectively).
On the PMT image (which matches the expected image), the first two beads appear to be distant of $\sim$3~\textmu m while they actually are $\sim$7~\textmu m apart. (Note that the third bead is below the background level.) From such image, the two beads could be either ~3, 5, 7 or 8~\textmu m apart. However, this ambiguity could be easily removed by slightly translating the sample while observing the image, in $x$ and $y$: if a bead is in the $\Delta \rr_a$ quadrant, it shifts in the same direction as the sample in both $x$ and $y$; in the $\Delta \rr_a^\pm$ quadrant, it shifts in the sample direction in $x$ and the opposite direction in $y$; etc.

In conclusion, we have demonstrated that, after optimizing the wavefront at the entrance of a SqMMF using the feedback from a single guide-star, four foci can be scanned at the fiber output by simply translating the input wavefront. This property originates from the previously observed memory effect in SqMMFs. We have shown that this effect can be exploited for fluorescence imaging in the vicinity of the guide-star, an approach that only requires access to the fiber proximal end. Although the persistence of the memory effect presently limits the field-of-view to $\sim$10$\times$10~\textmu m$^2$, this area remains stable across most of the fiber section, which is not the case for circular fibers. We anticipate that by combining \textit{a priori} models about the transmission of a SqMMF with information provided by a guide-star, imaging fidelity and field-of-view could be improved, making SqMMFs a promising alternative for micro-endoscopy.

\textbf{Funding} H2020 European Research Council (COHERENCE 681514); H2020 Marie Sklodowska-Curie Actions (DARWIN 750420).

\textbf{Acknowledgments} The authors thank Dorian Bouchet and Antonio Caravaca-Aguirre for fruitful discussions and insightful advice.

\end{document}